\documentclass[prl,twocolumn,aps,floatfix,longbibliography,nofootinbib]{revtex4-2}
\usepackage{pslatex,graphicx,dcolumn,bm,natbib,amssymb,amsmath,color,mathtools}
\usepackage[colorlinks=true,
            linkcolor=blue,       % internal refs: figures, equations, sections, tables
            citecolor=blue,       % citations (\cite)
            urlcolor=blue,        % URLs
            filecolor=blue,
            breaklinks=true]{hyperref}
\usepackage[utf8]{inputenc}
\usepackage[T1]{fontenc}
\usepackage{soul}
\usepackage{comment}
\usepackage{multirow}
\usepackage{xcolor}

\newcommand{\be}{\begin{equation}}
\newcommand{\ee}{\end{equation}}

\DeclarePairedDelimiter\abs{\lvert}{\rvert}

\newcommand{\ephi}{\hat{\mathbf e}_{\varphi, i}}
\newcommand{\ephio}{\hat{\mathbf e}_{\varphi, i}^0}

\begin{document}

%\title{Crowds spin against their members: chirality reversal in active rotor clusters}
%\title{Crowds spin against their members: Collective chirality reversal in active rotors}
%\title{Collective Chirality Reversal in Active Rotor Clusters}
\title{Emergent Chirality Reversal in Magnetically Driven Rotor Clusters}

\author{Xinzhi Li$^{1}$, Ke Zhong$^{2}$, Mohammad F. Islam$^2$, Shiladitya Banerjee$^1$}
\email{Correspondence: sbanerjee347@gatech.edu}
\affiliation{$^1$School of Physics, Georgia Institute of Technology, Atlanta, Georgia 30332, USA \\
$^2$Department of Materials Science and Engineering, Carnegie Mellon University, Pittsburgh, PA 15213, USA} 

\begin{abstract}
Does a collection of magnetically driven rotors necessarily rotate in the same direction as its individual members? We show that rotors driven by a rotating magnetic field self-assemble into clusters whose collective rotation can oppose the self-rotation of every rotor within them. The reversal arises from competition between rotation–translation coupling and dipolar interactions, which favor opposite directions of collective rotation. Their balance varies with system size and driving frequency, producing a transition between the two handed states, while hydrodynamic interactions shift the transition boundary. Collective handedness can thus emerge opposite to the field-imposed single-rotor rotation without reversing the external field, and can be tuned through system size and driving frequency.

\end{abstract}
\raggedbottom
\maketitle
Systems driven far from equilibrium can exhibit collective states with broken symmetries that have no equilibrium counterpart. Chirality provides a particularly striking example, arising across systems ranging from living matter to externally driven synthetic materials. In biological systems, chiral collective behavior appears in self-organized patterns of cells and bacteria~\cite{Ben-Jacob_PRL_1995, Tee_Nat_Cell_Bio_2015, Riedel_Science_2005}, asymmetric morphogenesis~\cite{Naganathan_Elife_2014,Smith_AnnuRev_2019}, and the collective chiral motion of active organisms~\cite{Petroff_PRL_2015, Beppu_PNAS_2021, Wu_PNAS_2011, Tan_Nature_2022}. In synthetic systems, externally driven particles provide a controllable setting in which chirality can be imposed and manipulated through driving protocols such as rotating fields~\cite{Xie_Sci_Rob_2019,Liu_PNAS_2020,Wang_PRL_2024,Chen_NatPhys_2025}. Such systems have also enabled applications in transport, sensing, microrobotics, optical materials, bioengineering, and environmental monitoring~\cite{Xie_Sci_Rob_2019,Gardi_Nat_Commun_2022,Chan_Nat_Commun_2024,Siebers_Sci_Adv_2023,Ai_Soft_Matter_2015,Lin_Mater_Today_2019,Wang_Sci_Adv_2021,Chan_Nat_Commun_2024}.

When chirality is imposed at the level of individual driven particles, the natural expectation is that the collective motion inherits the handedness of its constituents, as occurs in many driven and living systems~\cite{Petroff_PRL_2015,Xie_Sci_Rob_2019,Beppu_PNAS_2021,Tan_Nature_2022}. 
%Many-body interactions, however, can reorganize the dynamics of the constituent particles and need not preserve this correspondence. 
Whether a collection of identically chiral units can develop collective motion of the opposite handedness, and what controls such a reversal, remains unknown.

Here we investigate a minimal model of rotors driven by a rotating magnetic field. Each rotor carries an in-plane permanent moment that aligns with the field, resulting in self-rotation with a handedness set by the external drive. Through rotation–translation coupling, this self-rotation further generates a translational drive about the center of rotation of the magnetic field. Dipolar interactions drive the rotors to self-assemble into compact, ordered clusters without an imposed trapping potential. These clusters can rotate collectively in the direction opposite to the self-rotation of their constituent rotors, with the reversal controlled by the system size $N$ and the magnitude of the driving frequency $\omega$. We show that the reversal arises from competition between the rotation–translation drive, which favors counterclockwise collective rotation, and dipolar-induced co-rotation, which favors clockwise collective rotation through the structural anisotropy of the assembled cluster. The reversal follows from how these contributions scale: the rotation-translation torque grows
with cluster size faster than the dipolar torque, and the dipolar co-rotation weakens as the driving
frequency increases. Hydrodynamic interactions provide an additional contribution to clockwise co-rotation and shift the transition boundary but are not required for the reversal. 

\textit{Model.\textemdash}
We model the magnetic rotors in two dimensions using coupled rotational and  translational dynamics~\cite{Grzybowski_PRE_2001,Grzybowski_PNAS_2002,Zhong_Thesis_2024}.
%The surrounding fluid is treated in the low-Reynolds-number limit, so that hydrodynamic interactions are governed by the Stokes equation, while particle inertia is retained.
Each rotor is described by its position $\mathbf{r}_i$ and phase $\theta_i$, representing the orientation of its in-plane magnetic moment. The rotational dynamics follow the torque balance equation:
\begin{equation}
I \frac{{\rm d}^2 \theta_i}{{\rm d}t^2}
=
-\alpha \frac{{\rm d}\theta_i}{{\rm d}t}
+
MB \sin(\omega t - \theta_i),
\end{equation}
where $I$ is the moment of inertia, $\alpha$ is the rotational viscous drag coefficient, $M$ is the magnetic dipole moment of the rotor, and $B$ is the magnitude of the externally applied rotating magnetic field. This equation describes the alignment of the magnetic moment with a field rotating at angular frequency $\omega$.

The translational motion obeys the force balance equation:
\begin{equation}
m \frac{{\rm d}\mathbf{v}_i}{{\rm d}t}
=
-\beta \mathbf{v}_i
+
\mathbf{F}_i^{\rm rt}
+
\mathbf{F}_i^{\rm hyd}
+
\mathbf{F}_i^{\rm dip}
+
\mathbf{F}_i^{\rm rep}.
\label{eq:translation}
\end{equation}
The drag term $-\beta \mathbf{v}_i$ represents viscous resistance from the surrounding fluid, while the particle mass $m$ sets a finite relaxation timescale for the translational motion.
The rotation--translation drive $\mathbf{F}_i^{\rm rt}$ represents an effective coupling between rotor self-rotation and translational motion~\cite{Happel_1983,Pedrero_Sci_Adv_2018}. For a rotor
spinning near a substrate, a small tilt of the local field axis away from the vertical, which grows with distance from the field center, converts self-rotation into rolling along the azimuthal direction about the field center~\cite{SupInfo}. We model this coupling as
$
\mathbf{F}_i^{\rm rt}
=
-\gamma r_0^2\dot{\theta}_i\,\ephio,
$
where $\gamma$ is the rotation--translation coupling coefficient, $r_0$ is the particle radius, and $\ephio$ is the instantaneous azimuthal unit vector defined about the center of the externally applied rotating magnetic field. Unless stated otherwise, rotors are initialized in a disk centered on the field axis. Hydrodynamic interactions $\mathbf{F}_i^{\rm hyd}$ are modeled by the rotlet flow of a rotating particle in Stokes flow. The rotation of particle $j$ generates a viscous flow field that advects particle $i$~\cite{Blake_1974,Grzybowski_PRE_2001,Grzybowski_PNAS_2002,Pedrero_Sci_Adv_2018}. The resulting hydrodynamic force is
$
\mathbf{F}_i^{\rm hyd}
=
A
\sum_{j\neq i}
\dot{\theta}_j
\frac{
\hat{\mathbf z}\times\mathbf r_{ij}
}{
r_{ij}^{3}
},
$
where $\mathbf r_{ij}=\mathbf r_i-\mathbf r_j$, $r_{ij}=\abs{\mathbf r_{ij}}$, and $A$ sets the hydrodynamic coupling strength. Since
$\abs{\hat{\mathbf z}\times\mathbf r_{ij}}=r_{ij}$,
the magnitude of the pairwise interaction decays as $r_{ij}^{-2}$.
Dipolar interactions $\mathbf{F}_i^{\rm dip}$ arise from the magnetic dipole--dipole coupling between the rotors. Under the rotating field, the magnetic moments become aligned, leading to an effective attraction that promotes cluster formation (see Supplemental Information~\cite{SupInfo} for details). 
Finally, particle overlap is prevented by a short-range repulsive force,
$
\mathbf{F}_i^{\rm rep}
=
\sum_{j\neq i}
K_{\rm rep}\,(2 r_0 -r_{ij})_+
\,\hat{\mathbf r}_{ij},
$
where $K_{\rm rep}$ is the repulsive interaction coefficient, and $(x)_+=\max(x,0)$ ensures that the force acts only when two particles overlap.

\begin{figure}[t]
\begin{center}
\includegraphics[width=1\columnwidth]{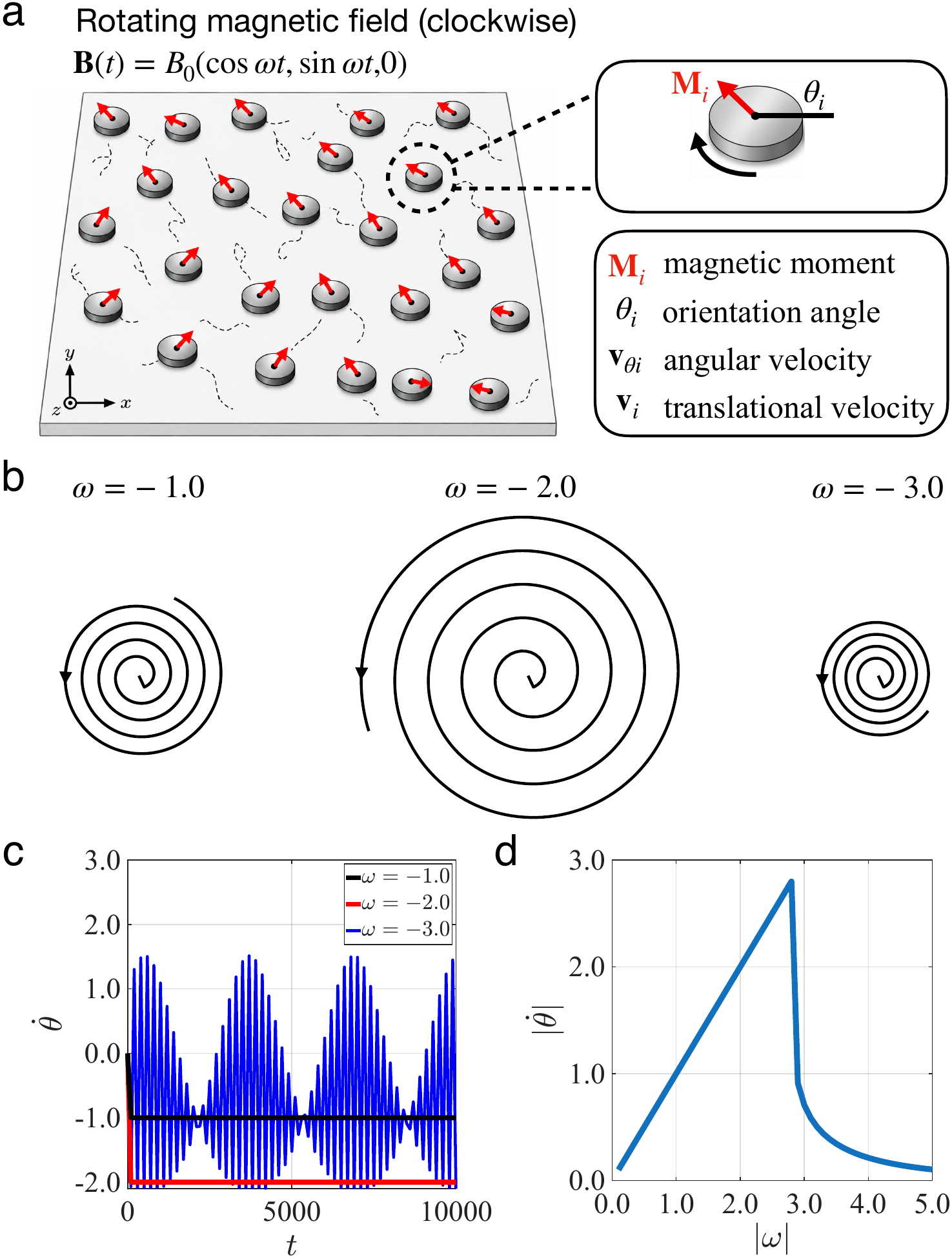}
\caption{\textbf{Single-particle dynamics and microscopic chirality.}
{\bf (a)} Schematic of the magnetically driven rotor model. Rotors with in-plane magnetic moments are driven by a rotating magnetic field $\mathbf{B}(t)=B_0(\cos\omega t,\sin\omega t,0)$, which induces self-rotation and translational motion.
{\bf (b)} Representative trajectories of isolated rotors for different driving frequencies $\omega$, showing circulation about the field center with a sense set by the field.
{\bf (c)} Time evolution of the angular velocity $\dot{\theta}$ for different driving frequencies. Increasing $\abs{\omega}$ leads to a crossover from synchronized rotation to oscillatory rotational dynamics.
{\bf (d)} Mean self-rotation angular velocity as a function of driving frequency, showing the transition between phase-locked and oscillatory regimes.
}
\label{fig1}
\end{center}
\end{figure}

All equations are nondimensionalized using the particle radius $r_0$ as the characteristic length scale and the magnetic alignment time $\omega_c^{-1}=\alpha/MB$ as the characteristic time scale. The resulting dimensionless model is governed by parameters characterizing viscous drag, hydrodynamic and dipolar interactions, steric repulsion, and the driving frequency~\cite{SupInfo}. The coupled equations are integrated numerically using an explicit time-stepping scheme from random initial configurations until a steady collective state is reached. By systematically varying the system size $N$ and driving frequency $\omega$, we characterize rotor self-assembly, emergence of collective motion and the transition between clockwise and counterclockwise rotation. %Full details of the model, parameter definitions, and an estimate of the Reynolds number are provided in the Supplemental Information~\cite{SupInfo}.

\textit{Single-particle dynamics and microscopic chirality.\textemdash}
We begin by characterizing the dynamics of an isolated rotor driven by a rotating external magnetic field, as illustrated in Fig.~\ref{fig1}(a). 
Each rotor possesses an in-plane magnetic moment $\mathbf m$ that tends to align with the rotating magnetic field $\mathbf B(t)=B_0(\cos\omega t,\sin\omega t)$. The resulting magnetic torque drives self-rotation of the particle, while the rotation--translation coupling of strength $\gamma$ converts self-rotation into a drive tangential to circles about the field center. The coupled dynamics therefore generates both spinning and translational motion at the single-particle level.
The resulting trajectories are shown in Fig.~\ref{fig1}(b) and Movie~S1. The particle circulates about the field center along slowly expanding spirals. The drive is purely azimuthal and no restoring force acts, so particle inertia produces a slow outward drift and the orbit radius grows with time. The size of the spiral reflects the mean spin rate, which is largest near the synchronization threshold [compare Fig.~\ref{fig1}(b) with Fig.~\ref{fig1}(d)]. The orbital angular velocity, of order $\gamma|\dot\theta|/(\beta r)$, is much smaller than the self-rotation rate.

To analyze the rotational dynamics, we introduce the phase difference
$\delta=\omega t - \theta$,
which obeys
$
\ddot{\delta}+\alpha \dot{\delta}
=
\alpha \omega-\xi \sin \delta .
$ Here $\alpha$ and $\xi$ are the dimensionless rotational drag and magnetic torque coefficients, respectively.
A phase-locked state corresponds to $\dot{\delta}=0$, yielding
$
\sin \delta = \alpha \omega / \xi,
$
which exists only for
$\abs{\omega}\le \xi / \alpha$. Because rotor inertia is appreciable for the parameters used ($\alpha^2\ll\xi$), locked and running states coexist over a range of driving frequencies, and a rotor started from rest loses synchronization at $\abs{\omega}\approx(4/\pi)\sqrt{\xi}\approx2.8$ [Fig.~\ref{fig1}(d)], well below $\xi/\alpha$.

As shown in Fig.~\ref{fig1}(c), the rotor exhibits a crossover from phase-locked motion at low driving frequencies, characterized by steady rotation, to an oscillatory regime at higher frequencies where the instantaneous angular velocity undergoes repeated sign reversals. Despite this, a finite mean rotation persists due to incomplete phase slipping. This is reflected in Fig.~\ref{fig1}(d), which shows that the mean self-rotation angular velocity deviates from the driving frequency as oscillations become pronounced, marking the breakdown of synchronization.

Because translational relaxation is slow compared with the spin, the orbital drive responds to the mean spin rate, which retains the sign of $\omega$ even in the running regime. The senses of both motions are therefore fixed by the field: for clockwise driving ($\omega<0$), every isolated rotor spins clockwise and circulates counterclockwise about the field center. Which of these two field-imposed senses a cluster adopts is decided by interactions, as shown below.

\begin{figure}[t!]
\begin{center}
\includegraphics[width=1\columnwidth]{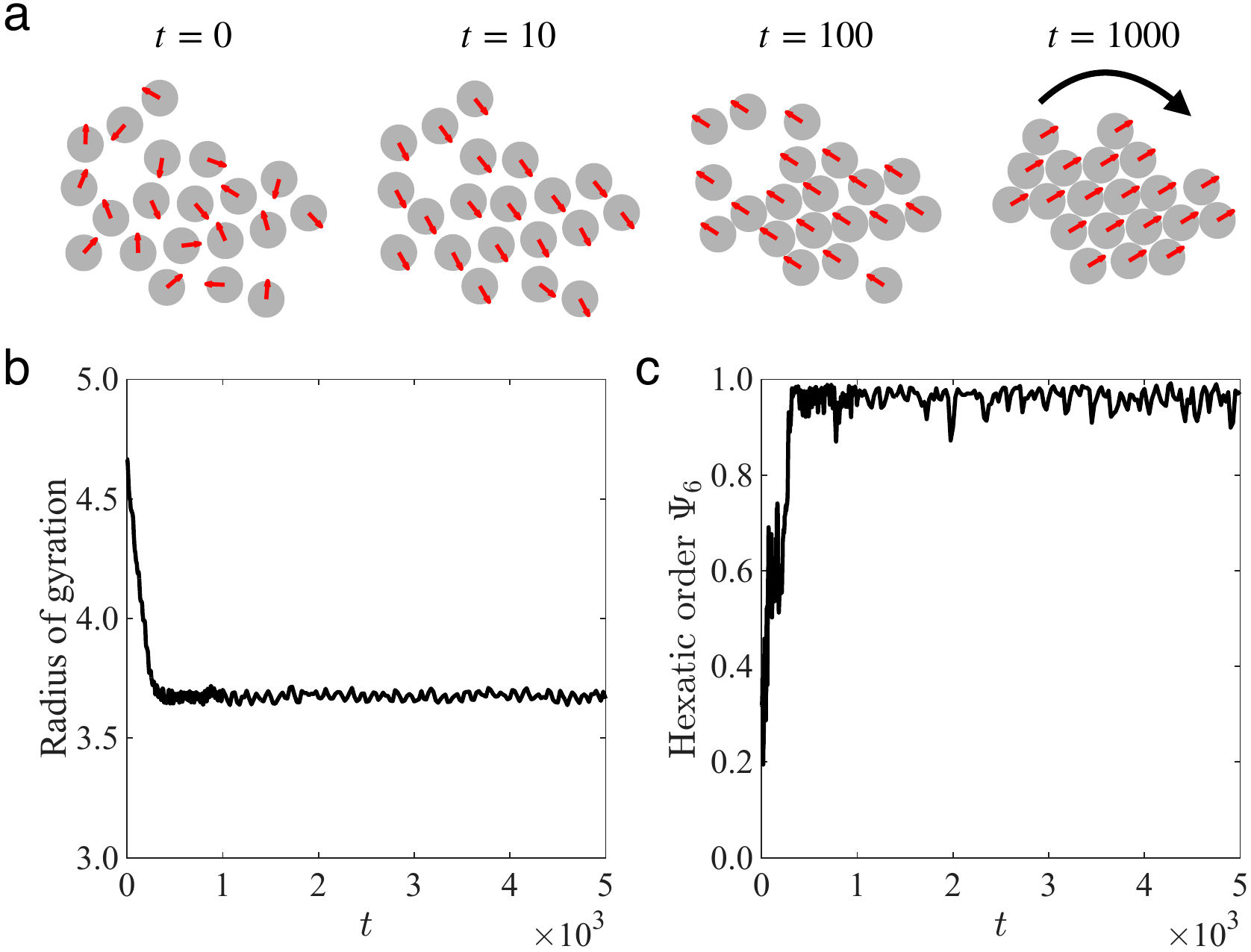}
\caption{
\textbf{Interaction-driven self-assembly of magnetically driven rotors.}
{\bf (a)} Representative snapshots showing that initially disordered particles aggregate into a compact ordered cluster in the absence of external confinement. The assembled rotor cluster also exhibits clockwise collective rotation as indicated by the black arrow.
{\bf (b)} Time evolution of the radius of gyration $R_g$, showing cluster compaction during assembly.
{\bf (c)} Evolution of the hexatic order parameter $\Psi_6$, indicating the emergence of local structural order.
}
\label{fig2}
\end{center}
\end{figure}

\textit{Interaction-induced self-assembly.\textemdash}
We next turn to the collective behavior of interacting rotors, which exhibit self-assembly even in the absence of external confinement or central attraction. In Fig.~\ref{fig2}(a), the system evolves from initially dispersed rotors into compact clustered states  (Movies~S2 and S3). Aggregation results from the period-averaged dipolar attraction between co-rotating moments. Because the moments rotate in phase, the pair interaction averaged over one field period is isotropic and attractive.

We characterize the assembly process through the cluster radius of gyration $R_g$. As shown in Fig.~\ref{fig2}(b), $R_g$ rapidly decreases during the initial stage of evolution and subsequently saturates, indicating the formation of a compact and dynamically stable aggregate. The assembled clusters further develop pronounced local structural order, characterized by the hexatic order parameter
$
\psi_6=
\abs*{
\frac{1}{N}
\sum_{i=1}^{N}
\frac{1}{N_i}
\sum_{j\in N_i}
e^{6 i \theta_{ij}}
},
$
where $\theta_{ij}$ denotes the bond angle between neighboring particles and $N_i$ is the coordination number of particle $i$. Fig.~\ref{fig2}(c) shows that $\psi_6$ rapidly approaches a value close to unity and remains large at long times, demonstrating the emergence of strong local six-fold ordering within the assembled cluster. 

\begin{figure}[t]
\begin{center}
\includegraphics[width=1\columnwidth]{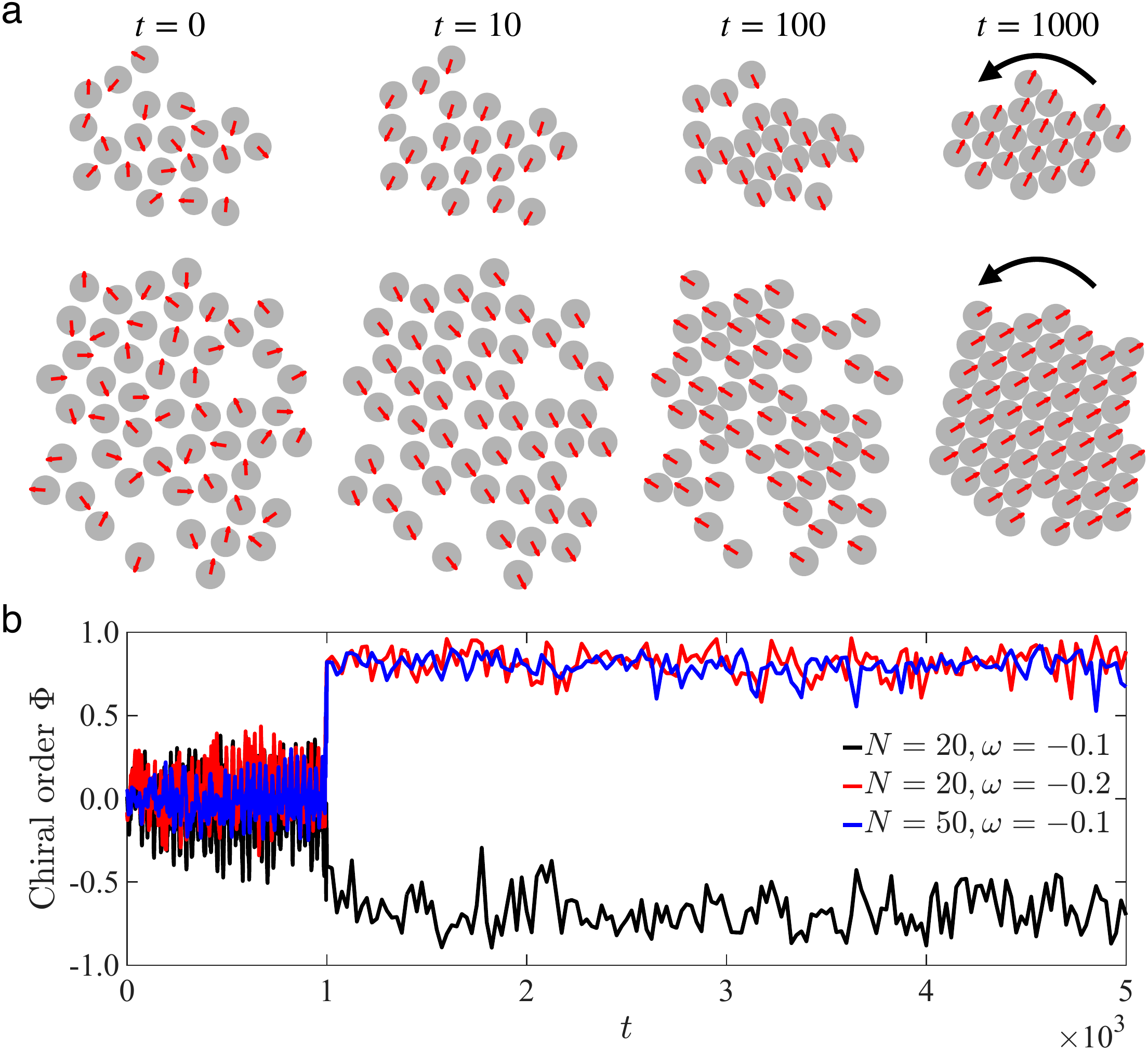}
\caption{
\textbf{Emergent chirality reversal in self-assembled rotor clusters.}
\textbf{(a)} Time-lapse snapshots illustrating two representative clusters exhibiting CCW collective rotation, opposite to the CW rotation shown in Fig.\ref{fig2}.
Top: $N=20, \omega=-0.2$.
Bottom: $N=50, \omega=-0.1$.
%Snapshots are shown at $t=0, 10, 100$, and $1000$.
\textbf{(b)}  Evolution of the chiral order parameter $\Phi$ for representative system sizes and driving frequencies. The red and blue curves are associated with the trajectories shown in panel (a), while the black curve represents the state shown in Fig.~\ref{fig2}.
}
\label{fig3}
\end{center}
\end{figure}

\textit{Emergent collective chirality and chirality reversal.\textemdash}
The self-assembled rotor clusters exhibit coherent collective rotation whose direction depends sensitively on the system size and driving frequency. Although every rotor spins clockwise at the rate imposed by the field, the assembled clusters can rotate either clockwise (CW) or counterclockwise (CCW), demonstrating a reversal of chirality as an emergent many-body phenomenon. Representative CW and CCW rotating states are presented in Fig.~\ref{fig2}(a) and Fig.~\ref{fig3}(a), with the corresponding dynamics shown in Movies~S2--S3 and S4--S7, respectively.

The chirality of the collective rotation is characterized by the chiral order parameter~\cite{Xie_Sci_Rob_2019,Liu_Nature_2021}
$
\Phi(t)
=
\frac{1}{N}
\sum_{i=1}^{N}
\ephi
\cdot
\frac{\mathbf v_i(t)}
{\abs{\mathbf v_i(t)}},
$
where $\ephi$ is the counterclockwise azimuthal unit vector about the instantaneous cluster centroid and $\mathbf v_i$ is the velocity of particle $i$. By construction, $\Phi>0$ ($\Phi<0$) corresponds to counterclockwise (clockwise) collective rotation, respectively. Starting from a disordered initial state, $\Phi(t)$ evolves through a transient self-assembly stage before converging to a steady value of either sign, indicating robust CCW or CW collective rotation [Fig.~\ref{fig3}(b)]. The frequency of the collective rotation is quantified separately by the cluster angular velocity $\omega_{\rm cluster}$, whose definition and dependence on system parameters are presented in SI Sec.~III and Figs.~S7-S8.

\begin{figure}[t]
\begin{center}
\includegraphics[width=1\columnwidth]{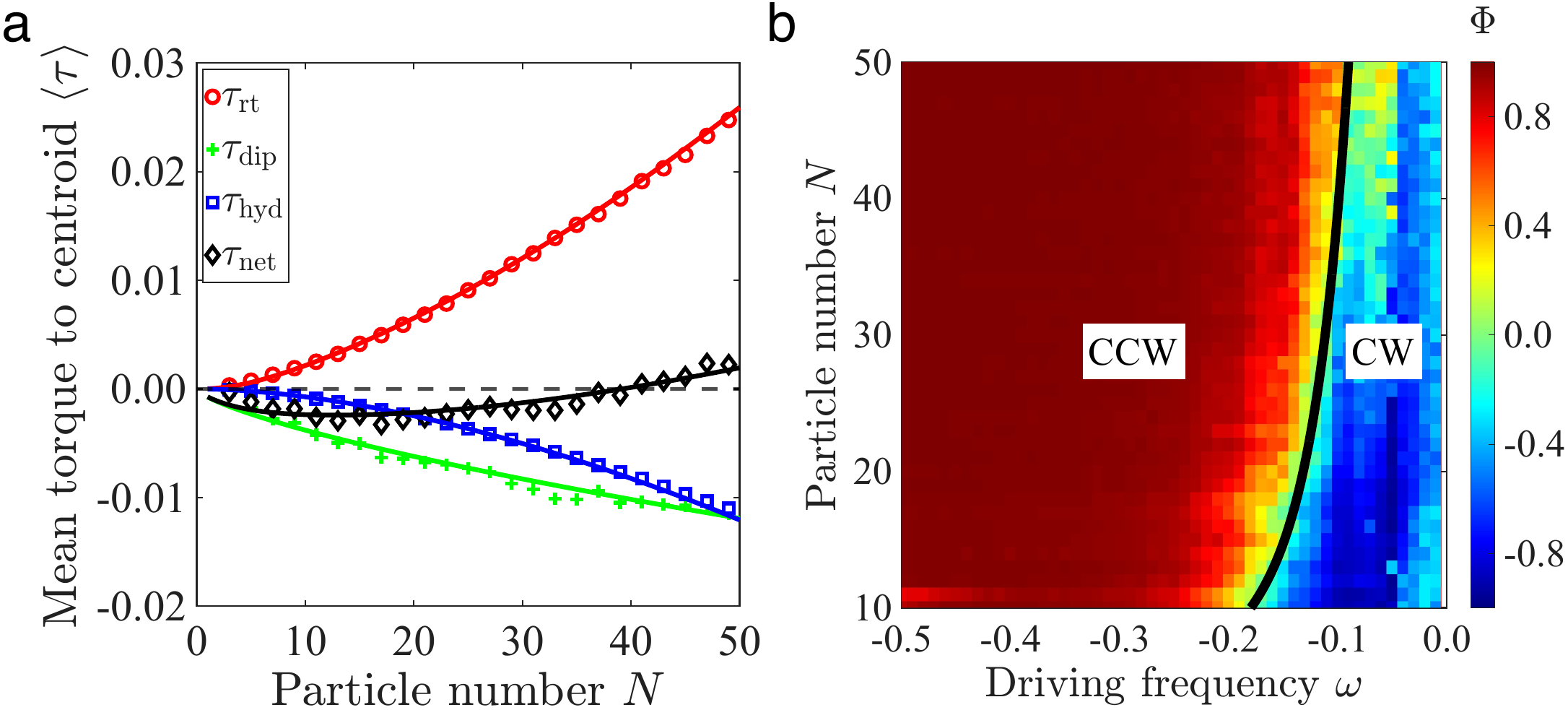}
\caption{
\textbf{Torque competition underlying collective chirality reversal.}
\textbf{(a)} Mean torque about the cluster centroid as a function of particle number $N$ at $\omega=-0.1$. The rotation--translation torque $\tau_{\rm rt}$ favors CCW collective rotation, whereas the dipolar torque $\tau_{\rm dip}$ and hydrodynamic torque $\tau_{\rm hyd}$ favor CW rotation. Symbols show the torques measured directly from the microscopic forces, while solid curves show the corresponding fitted torque expressions. The resulting net torque $\tau_{\rm net}=\tau_{\rm rt}+\tau_{\rm dip}+\tau_{\rm hyd}$ changes sign as the rotation--translation contribution overcomes the opposing dipolar and hydrodynamic torques.
\textbf{(b)} Steady-state chiral order parameter $\Phi$ as a function of particle number $N$ and driving frequency $\omega$. Positive (negative) values indicate CCW (CW) collective rotation. The solid black curve denotes the zero-net-torque boundary calculated from the fitted torque expressions, which captures the observed chirality-reversal boundary.
}
\label{fig4}
\end{center}
\end{figure}

The steady-state chirality over a broad range of system sizes and driving frequencies is summarized in the phase diagram in Fig.~\ref{fig4}(b). For small clusters, the collective rotation follows the CW self-rotation of the individual rotors. As the cluster size increases, however, the system crosses into a CCW state, demonstrating a size-induced reversal of collective chirality. The location of this transition also depends systematically on the driving frequency, with the critical cluster size decreasing as $\abs{\omega}$ increases. These results reveal that the collective handedness is controlled jointly by system size and external driving, even though the self-rotation of every rotor is fixed by the field.

\textit{Mean-field theory of chirality reversal.\textemdash}
The physical origin of the chirality reversal can be understood through a mean-field description of the collective rotational dynamics, as developed in the Supplemental Information~\cite{SupInfo}. The rotation--translation drive generated by the field-imposed self-rotation of the individual rotors favors CCW
collective rotation, whereas dipolar interactions favor CW co-rotation through the coupling between the rotating magnetic moments and the structural anisotropy of the cluster, and hydrodynamic advection by neighboring rotors adds a further CW contribution.
%whereas the dipolar and hydrodynamic interactions favor CW rotation through the coupling between the rotating magnetic moments and the structural anisotropy of the cluster. 
Their competition determines the direction of collective rotation, which can reverse as the particle number or driving frequency is varied. Taking the moment of Eq.~\eqref{eq:translation} about the centroid gives an exact equation for the rotational moment of the cluster, $L_c=\sum_i[\mathbf s_i\times(\mathbf v_i-\mathbf V_{\rm cm})]\cdot\hat{\mathbf z}$, where $\mathbf s_i$ is the position of rotor $i$ relative to the centroid: $\dot L_c+\beta L_c=\tau_{\rm rt}+\tau_{\rm hyd}+\tau_{\rm dip}$. Steric forces are central and drop out. In the steady state, the sign of the mean collective rotation is therefore the sign of the mean net torque. 

To quantify this competition, we compute the torque exerted by each interaction about the cluster centroid. The rotation--translation, hydrodynamic, and dipolar contributions can be written as
\begin{equation}
\tau_{\rm rt}=\gamma|\omega|N\bar{s}\Lambda_{\rm rt},\,
\tau_{\rm hyd}=-A|\omega|N\bar{s}\Lambda_{\rm hyd},\,
\tau_{\rm dip}=K_{\rm dip}\mathcal{A}\sin2\chi .
\label{eq:torque_components}
\end{equation}
Here $\bar{s}$ is the mean distance of the particles from the cluster centroid, $\Lambda_{\rm rt}$ and $\Lambda_{\rm hyd}$ quantify the geometric coupling of the corresponding interactions to collective rotation, $\mathcal{A}$ measures the two-fold structural anisotropy of the cluster, and $\chi$ is the angular lag between the magnetic moments and the structural easy axis. These expressions reproduce the torques measured directly from the microscopic forces. The dipolar expression is exact when all moments are parallel.
%These reduced expressions quantitatively reproduce the torques measured directly from the microscopic forces. 

To obtain a closed description, we represent the particle-number and frequency dependence of the structural quantities in Eq.~\eqref{eq:torque_components} by fits to the simulation data (SI Sec.~SII and Figs.~S1--S3). Because the torque balance is exact, the boundary obtained in this way is not a parameter-free prediction. Its value lies in identifying which contributions control the reversal and how they scale. Figure~\ref{fig4}(a) compares the resulting fitted torque expressions with the torques measured directly from the microscopic forces. The rotation-translation contribution favors CCW rotation, whereas the dipolar and hydrodynamic contributions favor CW rotation. With increasing particle number or driving-frequency magnitude, the rotation--translation contribution becomes sufficiently strong to overcome the opposing torques. The fits make the reason explicit. For a compact cluster, $\bar s\propto N^{1/2}$ and $\Lambda_{\rm rt}$ saturates, so $\tau_{\rm rt}\propto\abs{\omega}N^{3/2}$, whereas the dipolar torque grows only as $N^{0.71}$ (SI Fig.~S3), consistent with cancellation of the two-fold anisotropy in the sixfold-ordered interior, which leaves mainly the cluster boundary to contribute. The hydrodynamic torque, $\tau_{\rm hyd}\propto\abs{\omega}N^{3/2}\ln N$, grows marginally faster than $\tau_{\rm rt}$ but remains smaller over the simulated range because $\gamma>A$. Increasing $\abs{\omega}$ also favors the CCW drive: $\tau_{\rm rt}$ and $\tau_{\rm hyd}$ grow linearly with $\abs{\omega}$, while the dipolar torque decreases, consistent with the cluster increasingly slipping behind the rotating moments. 

In the statistically stationary state, chirality reversal occurs when the mean net torque changes sign. The phase boundary is therefore givend by
\begin{equation}
\left\langle\tau_{\rm net}\right\rangle_{\rm fit}
=
\tau_{\rm rt}^{\rm fit}
+
\tau_{\rm hyd}^{\rm fit}
+
\tau_{\rm dip}^{\rm fit}
=0 .
\label{eq:torque_balance_boundary}
\end{equation}
which closely follows the CW--CCW transition in Fig.~\ref{fig4}(b). Hydrodynamic interactions provide an additional CW contribution and shift the reversal boundary against the CCW rotation--translation drive, but are not required for the reversal: when hydrodynamic interactions are removed, the measured rotation--translation and dipolar torques alone exhibit a zero-net-torque boundary that follows the chirality transition (SI Fig.~S4).

\textit{Discussion.\textemdash}
Our results identify a minimal mechanism by which collective handedness can differ from the field-imposed handedness of the individual constituents. Previous studies have shown that chiral motion can originate from particle shape, surface asymmetry, or assembled chiral structures~\cite{Kummel_PRL_2013, Brooks_Nat_Commun_2019, Mcneill_ACS_Nano_2023, Nourhani_PhysRevE_2016, Wykes_Soft_Matter_2016, Aubret_Nat_Phys_2018, Schmidt_J_Chem_Phys_2019}. Here, the constituent particles are isotropic and share the same rotational handedness, yet the interplay between local rotation--translation coupling and collective dipolar interactions allows the assembled cluster to rotate with either handedness. More generally, whenever local chiral driving competes with collective interactions that favor opposite handedness and respond differently to system size or driving conditions, the balance can reverse the chirality at the collective scale. Related competition may arise in synthetic active systems~\cite{Chen_NatPhys_2025} and in biological systems, including assemblies of spinning organisms~\cite{Tan_Nature_2022}, active cytoskeletal systems~\cite{Naganathan_Elife_2014}, and cilia-driven flows~\cite{Smith_AnnuRev_2019}.

The mechanism makes predictions that can be tested with magnetic rotors driven near a substrate or at a liquid interface~\cite{Grzybowski_PRE_2001,Grzybowski_PNAS_2002,Pedrero_Sci_Adv_2018}. At fixed field, the sense of collective rotation should switch with the number of rotors in a cluster alone, and the reversal boundary should depend on the position of the cluster relative to the field axis (SI Fig.~S6). Our model is two dimensional, represents the rotation--translation drive phenomenologically, and neglects wall corrections to the hydrodynamic interaction; how these simplifications, together with confinement and spatially varying driving, affect the reversal remains to be established.

\bibliography{chiral_materials.bib}

\end{document}